# Excess-Hole Induced High Temperature Polarized State and its Correlation with the Multiferroicity in Single Crystalline DyMnO$_3$

Tao Zou,[1)] Zhiling Dun,[2)] Huibo Cao,[3)] Mengze Zhu,[1)] Daniel Coulter,[1)] Haidong Zhou,[2)] Xianglin Ke[1,a)]

[1]*Department of Physics and Astronomy, Michigan State University, East Lansing, MI 48824, USA*

[2]*Department of Physics and Astronomy, University of Tennessee, Knoxville, TN 37996, USA*

[3]*Quantum Condensed Matter Division, Oak Ridge National Laboratory, Oak Ridge, TN 37831, USA*

Controlling the ferroelectricity and magnetism in multiferroic materials has been an important research topic. We report the formation of a highly polarized state in multiferroic DyMnO$_3$ single crystals which develops well above the magnetic transition temperatures, and we attribute it the thermally stimulated depolarization current effect of excess holes forming Mn$^{4+}$ ions in the system. We also show that this high temperature polarized state intimately correlates with the lower temperature ferroelectric state that is induced by the incommensurate spiral magnetic order of Mn spins. This study demonstrate an efficient approach to tune the multiferroicity in the manganite system.

[a)] Author to whom correspondence should be addressed. Electronic mail: ke@pa.msu.edu

Multiferroics with strong magnetoelectric (ME) coupling effects have been widely studied in the past decade for both fundamental understanding of the underlying mechanisms and potential technology applications.[1-8] Many efforts have been devoted to tuning the ferroelectric polarization by an external magnetic field or modifying the magnetic ordering by an applied electric field.[5, 9] Along this line, via the electric field generated by a charge-trapping layer or ferroelectric layer in multiferroic composites, one could dramatically alter materials electrical transport or magnetic properties.[10-12]

Among the single-phase improper multiferroics in which the ferroelectricity is induced by the occurence of incommensurate (ICM) magnetic structures (i.e., spiral spin structures) below the magnetic transition, $RMnO_3$ (R = Tb and Dy) perovskites have been well studied since Kimura *et al*. initially discovered strong ME coupling effects and large magneto-capacitance in this system.[1, 13, 14] Both compounds show a distorted orthorhombic perovskite structure and undergo sequential magnetic phase transitions, i.e., an antiferromagnetic transition at $T_N$ with an ICM sinusoidal spin structure of Mn spins directing along the *b*-axis, followed by a transition at $T_S$ with a cycloidal spiral spin structure in the *bc* plane then by the lower temperature magnetic order of rare earth ions.[15-17] The spontaneous ferroelectric (FE) polarization emerges at $T_S$, arising from the coupling of a spiral ordering of Mn spins to the lattice that can be expressed as $\vec{P} \sim \vec{e}_{i,j} \times (\vec{S}_i \times \vec{S}_j)$, with $\vec{S}_{i(j)}$ representing the spin direction at site *i(j)* and $\vec{e}_{i,j}$ denoting the unit vector pointing from site *i* to site *j*.[18-20] An applied magnetic field along the *b*-axis flops the spin rotation from the *bc* plane to the *ac* plane, resulting in switching P from the *c*-axis to the *a*-axis.[13, 14, 21-23] This indicates a large ME coupling that is also corroborated by the control of spin chirality via the polarity of an electric field.[24] In this paper, we report an approach to tune the multiferroicity in single crystalline multiferroic $DyMnO_3$ via an internal electric field exerted by trapped charges that are formed well above $T_s$ and antiferromagnetic transtion $T_N$. This is in contrast to the regular method with an electric poling field applied through the materials improper ferroelectric transtion as reported in early literature.

Single crystalline $DyMnO_3$ was grown using the optical floating zone technique in an Ar atmosphere. The samples measured were cut with the out-of-plane orientation along the *c*-axis, which was determined by the Laue x-ray diffraction method. The $DyMnO_3$ sample we use has a c-axis thickness of 0.38 mm. Specific heat of $DyMnO_3$ was measured using the Quantum Design Physical Property Measurement System (PPMS), and the capacitance measurements were conducted with a HP capacitance bridge (HP 4191A Impedance Analyzer). After the samples were cooled down under various circumstances as described in the figure captions, the electric field was removed and the electrodes were shorted for long enough time prior to measuring the pyroelectric current with a Keithley 6517B electrometer. The sample was heated at a rate of b = 3 K/min, except for the heating rate dependence measurements shown in Fig. 2(d). The ferroelectric polarization was acquired by integrating the pyroelectric current as a function of time. The electrode contacts were made of silver paint, and a home-made sample probe incorporated with the PPMS was used for both capacitance and pyroelectric current measurements. Note all the pyroelectric current and capacitance data reported thereafter were measured with a poling field applied along the $c$-axis.

DyMnO$_3$ exhibits three anomalies in the low temperature specific heat measurement shown in Fig. 1(a) with $T_N$ ~ 39 K, $T_S$ ~ 18 K, and $T_{Dy}$ ~ 6.5 K, where $T_N$ and $T_S$ are associated with the ICM sinusoidal and spiral order of Mn spins, respectively, and $T_{Dy}$ is related to the ICM-commensurate magnetic transition of Dy spins.[13, 17] Correspondingly, it also shows two anomalies in the capacitance measurements (purple curve in Fig. 1a) corresponding to the FE transitions (Fig. 1b): one at $T_S$ and the other at $T_{Dy}$ at which a decrease in polarization (P) value is observed. Note that here an electric field of ± 487 kV / m (± 185 V voltage) was applied to pole the sample along the $c$-axis from $T$ = 25 K to 2 K and the pyroelectric current was integrated up to 50 K to obtain the P value. One can see that P is reversed with an opposite electric field and it approaches zero above $T_S$. These results are in good agreement with previous reports.[13, 17, 25]

The features we discovered in DyMnO$_3$ emerged when the sample was cooled down from 100 K with a poling voltage ($V_p$ = ± 185 V) and then the pyroelectric current was recorded up to 120 K shown in Fig. 1(c) and 1(d). In addition to those two sharp features occurring at $T_S$ and $T_{Dy}$ as seen in previous studies when poling the sample from a temperature just above $T_S$ or $T_N$,[1, 13] we found another pyroelectric current peak emerging at $T$ ~ 88 K. This additional peak is also reversible with an opposite electric field as shown in Fig. 1(c). The corresponding P curves are plotted in Fig. 1(d) where one can see that the P value contributed from the high temperature transition is ~ 0.2 μC/cm$^2$, even slightly larger than that induced by the magnetic order of Mn and Dy spins at $T_S$ and $T_{Dy}$, respectively.

We tried to identify the nature of this emergent state associated with the pyroelectric peak occuring above $T_N$, however, no anomalies in dielectric constant and heat capacity measurements were detected at $T$ ~ 88 K. And no structural phase transtition is observed above and below this temperature in both neutron and x-ray single crystal diffraction measurements. These facts suggest a non-ferroelectric character of the feature shown at $T \sim 88$ K. For convenience, we denote this emergent phase between $T_S \sim 18$ K and $T \sim 88$ K as High Temperature Polarized (HTP) state in the following discussion, and use the HTP polarization to describe the integration of pyroelectric current as a function of time within this temperature region.

Figure 2(a) displays the integrated P measured after the sample was cooled down from $T = 100$ K to 2 K with various $V_p$s. The magnitude of both FE and HTP polarization increase simultaneously with increasing $V_p$ (also seen in Fig. S1).[26] A sizable polarization is still present even with a $V_p$ as small as 10 V. We also fixed $V_p$ to + 185 V and cooled down the sample from various poling temperatures $T_{pole}$ to 2 K. The thus-obtained P data are shown in Fig. 2(b). We find that with decreasing $T_{pole}$ the HTP polarization is significantly suppressed especially when $T_{pole}$ is smaller than 100 K (shown in the inset of Fig. 2b) while keeping the low temperature FE state nearly intact. Remarkably, Fig. 2(b) shows a small negative P value for the HTP state when $T_{pole}$ is lower than 60 K. This originates from the negative pyroelectric current peak occurring at high temperature (see an expanded view shown in the bottom inset of Fig. 2c) which is about 2 orders of magnitude smaller than the pyroelectric current (see a full view shown in the top inset of Fig. 2c) at low temperatures such that it was highly likely to be overlooked in previous studies.[1, 13] Such a negative P behavior is also observed with different $V_p$s applied at $T_S < T < T_N$ (seen in Fig. S2) and is reversible with an opposite poling field as displayed in Fig. 2(c).[26] Naively, one would expect to observe a small positive P with a positive poling field. To understand the appearance of a negative P above $T_S$, one needs to take into account the correlation between FE and HTP states.

To explore the correlation between FE and HTP states, the sample was cooled down under different poling conditions to reach different polarized status of these two states. Fig. 3(a) displays polarization curves measured after the sample was cooled down from $T = 100$ K to 25 K with different $V_p$s (EC) and then zero-field cooled (ZEC) from 25 K to 2 K. The HTP polarization increases with increasing $V_p$ and has the same sign as $V_p$. Surprisingly, however, the net FE polarization contribution from $T_S > T > T_{Dy}$ is negative and its magnitude increases with the HTP polarization and saturates at high HTP polarization values (seen in Fig. 3b). Similar behavior was also observed when the sample was cooled from $T = 100$ K to various $T^*$ with 185 V voltage applied and then ZEC from $T^*$ to 2 K in Fig. S3.[26] Previous studies found that one can tune the spin chirality and thus the sign and magnitude of P by applying an electric poling field through $T_S$.[24] Therefore, without the correlation between FE and HTP states, one would expect the FE polarization is close to zero with zero-field cooling process from a temperature above $T_S$, if not exactly zero, due to the equal population of clockwise and counterclockwise spin chirality. This obviously contradicts with the experimental observation shown in Fig. 3(a), 3(b) and S3,[26] and thus overrules the aforementioned assumption, i.e., FE and HTP intimately correlate with each other.

As mentioned above, the HTP state could not be verified as ferroelectric even though the pyroelectric current is reversible. Instead, the HTP state can be ascribed to the thermally stimulated depolarization current (TSDC) of charge carriers.[27, 28] A recent study by Kohara has reported the existence of large pyroelectric current (and polarization) in yttrium iron garnet ($Y_3Fe_5O_{12}$) and attributed it to TSDC associated with excess electrons forming $Fe^{2+}$ ions.[28] In $DyMnO_3$, it is likely that the sample is nonstoichiometric with excess oxygen due to the growth conditions such that a small portion of Mn ions has a valence of 4+,[29] which provides excess hole carriers that are barely mobile at high temperature. When the sample is cooled down from 100 K under the poling electric field, these hole carriers in $DyMnO_3$ redistribute and form electric dipoles and thus an internal electric field ($E_{int}$) to screen the external field. The charge carriers are trapped at low temperature after retracting $V_p$ to zero due to the relatively long relaxation time. When warming up the sample, the trapped carriers are theremally released, leading to the pyroelectric current peak at high temperature. Fig. 2(d) shows the high temperature pyroelectric current measured with different warming rates, in which one can see that the pyroelectric current peak shifts to higher tempeature with increasing the warming rate, comfirming the thermal relaxation process of charge carriers that characterize the TSDC behavior in $DyMnO_3$. The sign of TSDC relative to that of the poling field is determined by the carrier type of trapped charges, i.e., electron or hole. For instance, the TSDC induced by the existence of trapped electrons in $Y_3Fe_5O_{12}$ has an opposite sign to the applied electric poling field.[28] But, for $DyMnO_3$ in this study the observed P and TSDC have the same sign as $V_p$, implying the existence of the excess hole carriers provided by $Mn^{4+}$ ions. This is affirmed by the decrease of the high temperature polarization value due to the reduction of oxygen content after the sample is annealed in Ar atmosphere at 900°C (Fig. S4). Thus, the induced $E_{int}$ has a direction opposite to $V_p$ and still exists after $V_p$ is removed, and $E_{int}$ combined with an additional external field is responsible for the low temperature FE state of DyMnO$_3$.

To further examine the correlation between FE and HTP states, schematics in Fig. 4 illustrate the low temperature magnetic structure and P direction under different cooling procedures. With ZFC from 100 K to 25 K, Mn$^{4+}$ ions are randomly distributed and become frozen. In this case, $E_{int}$ is negligible such that $V_p$ applied below 25 K will determine the spin chirality and thus P of the FE state has the same sign as $V_p$ (Fig. 4a, 1b, and S2).[26] On the other hand, when cooling down the sample with $V_p$ applied from 100 K to 25 K and then ZFC from 25 K to 2 K, the $E_{int}$ induced in the HTP state will determine the magnetic structure below $T_S$ and the associated FE state. Since $E_{int}$ has an opposite sign to $V_p$, the FE state has reversed spin chirality compared to the case shown in Fig. 4(a) and the induced P values in the FE and HTP states essentially have an opposite sign (Fig. 4b and 3a). Thus, one can see that $E_{int}$ induced in the HTP state and the external field contributed by $V_p$ compete with each other and such a relation is more explicitly demonstrated in Fig. 3(c) where the sample was cooled down from 100 K to 25 K with a fixed $V_p$ of + 185 V and then further cooled down to 2 K with various $V_p$s. Note that the total P value almost saturates at about $P_c$ ~ 0.2 µC/cm$^2$ down to 2 K with a $V_p$ of + 30 V, which implies no contribution to P from the FE state (Fig. 4c) and no preferable spin chirality below $T_S$. This suggests that the strength of the correlation between HTP (induced by + 185 V) and FE states can be expressed as an effective $E_{int}$ of - 79 kV/m applied through $T_S$ for the FE state. Other curves measured with various $V_p$s are nearly anti-symmetric about 0.2 µC/cm$^2$ at low temperatures. In other words, as plotted in Fig. 3(d), +30 V is a critical $V_p$ above which the net FE polarization is positive because of the dominant role of $V_p$ over $E_{int}$ and below which the net FE polarization is negative because of the dominant role of $E_{int}$ over $V_p$ or the combination of $E_{int}$ and $V_p$s with negative sign. For instance, with a single $V_p$ (+185 V) applied from from $T$ = 100 K to 2 K, the P values induced in both FE and HTP states are positive (Fig. 2a and 4d) and the low temperature FE state is different from the case shown in Fig. 3(a) and 4(b).

Finally, it is worth noting that the induced dipole in the FE state shown in Fig. 4(a) can in turn polarize the excess holes forming Mn$^{4+}$ ions to form a screening field with an opposite direction to P in the FE state, which results in small negative pyroelectric current (polarization) in the HTP state while warming, as seen in Fig. 2(c) and S2.[26]

To summarize, we have found a highly polarized state in single crystalline muliferroic DyMnO$_3$ above the magnetic phase transitions and attributed it to the TSDC effect of excess holes forming Mn$^{4+}$ ions. The induced internal electric field in such a high temperature polarized state intimately correlates with the low temperature FE state. This provides an efficient approach to tune the ferroelectric state as well as spin helicity in multiferroic materials.

**Acknowledgements**

X.K. acknowledges the support from the start-up funds at Michigan State University. Z. L. Dun and H.D.Z. thank for the support from NSF-DMR through award DMR-1350002. Work at ORNL was sponsored by the Scientific User Facilities Division, Office of Basic Energy Sciences, U.S. Department of Energy. T. Z. appreciates the useful discussion with Prof. J.M. Liu.

[1] T. Kimura, T. Goto, H. Shintani, K. Ishizaka, T. Arima, and Y. Tokura, Nature **426**, 55 (2003).

[2] M. Fiebig, J. Phys. D.: Appl. Phys. **38**, R123 (2005).

[3] S.-W. Cheong and M. i. Mostovoy, Nat. Mater. **6**, 13 (2007).

[4] J. Wang, J. B. Neaton, H. Zheng, V. Nagarajan, S. B. Ogale, B. Liu, D. Viehland, V. Vaithyanathan, D. G. Schlom, U. V. Waghmare, N. A. Spaldin, K. M. Rabe, M. Wuttig, and R. Ramesh, Science **299**, 1719 (2003).

[5] T. Zhao, A. Scholl, F. Zavaliche, K. Lee, M. Barry, A. Doran, M. P. Cruz, Y. H. Chu, C. Ederer, N. A. Spaldin, R. R. Das, D. M. Kim, S. H. Baek, C. B. Eom, and R. Ramesh, Nat. Mater. **5**, 823 (2006).

[6] Y. Tokura, Science **312**, 1481 (2006).

[7] N. A. Spaldin, S.-W. Cheong, and R. Ramesh, Phys. Today **63**, 38 (2010).

[8] K. F. Wang, J. M. Liu, and Z. F. Ren, Adv. Phys. **58**, 321 (2009).

[9] T. Nozaki, Y. Shiota, M. Shiraishi, T. Shinjo, and Y. Suzuki, Appl. Phys. Lett. **96**, 022506 (2010).

[10] U. Bauer, M. Przybylski, J. Kirschner, and G. S. Beach, Nano Lett. **12**, 1437 (2012).

[11] C. H. Ahn, M. Di Ventra, J. N. Eckstein, C. D. Frisbie, M. E. Gershenson, A. M. Goldman, I. H. Inoue, J. Mannhart, A. J. Millis, A. F. Morpurgo, D. Natelson, and J.-M. Triscone, Rev. Mod. Phys. **78**, 1185 (2006).

[12] C. A. Vaz, J. Phys.: Condens. Matter **24**, 333201 (2012).

[13] T. Goto, T. Kimura, G. Lawes, A. P. Ramirez, and Y. Tokura, Phys. Rev. Lett. **92**, 257201 (2004).

[14] T. Kimura, G. Lawes, T. Goto, Y. Tokura, and A. P. Ramirez, Phys. Rev. B **71**, 224425 (2005).

[15] M. Kenzelmann, A. B. Harris, S. Jonas, C. Broholm, J. Schefer, S. B. Kim, C. L. Zhang, S. W. Cheong, O. P. Vajk, and J. W. Lynn, Phys. Rev. Lett. **95**, 087206 (2005).

[16] T. Arima, A. Tokunaga, T. Goto, H. Kimura, Y. Noda, and Y. Tokura, Phys. Rev. Lett. **96**, 097202 (2006).

[17] O. Prokhnenko, R. Feyerherm, E. Dudzik, S. Landsgesell, N. Aliouane, L. C. Chapon, and D. N. Argyriou, Phys. Rev. Lett. **98**, 057206 (2007).

[18] M. Mostovoy, Phys. Rev. Lett. **96**, 067601 (2006).

[19] I. A. Sergienko and E. Dagotto, Phys. Rev. B **73**, 094434 (2006).

[20] H. Katsura, N. Nagaosa, and A. V. Balatsky, Phys. Rev. Lett. **95**, 057205 (2005).

[21] F. Kagawa, M. Mochizuki, Y. Onose, H. Murakawa, Y. Kaneko, N. Furukawa, and Y. Tokura, Phys. Rev. Lett. **102**, 057604 (2009).

[22]N. Aliouane, K. Schmalzl, D. Senff, A. Maljuk, K. Prokes, M. Braden, and D. N. Argyriou, Phys. Rev. Lett. **102**, 207205 (2009).

[23]N. Abe, K. Taniguchi, S. Ohtani, H. Umetsu, T. Arima, Phys. Rev. B **80**, 020402 (2009).

[24]Y. Yamasaki, H. Sagayama, T. Goto, M. Matsuura, K. Hirota, T. Arima, and Y. Tokura, Phys. Rev. Lett. **98**, 147204 (2007).

[25]E. V. Milov, A. M. Kadomtseva, G. P. Vorob'ev, Y. F. Popov, Y. Ivanov, A. Mukhin, and A. M. Balbashov, Jetp Lett. **85**, 503 (2007).

[26]See supplemental material at [URL will be inserted by AIP publishing] for more information.

[27]C. Bucci, R. Fieschi, and G. Guidi, Phys. Rev. **148**, 816 (1966).

[28]Y. Kohara, Y. Yamasaki, Y. Onose, and Y. Tokura, Phys. Rev. B **82**, 14419 (2010).

[29]C. Ritter, M. R. Ibarra, J. M. De Teresa, P. A. Algarabel, C. Marquina, J. Blasco, J. García, S. Oseroff, and S-W. Cheong, Phys. Rev. B **56**, 8902 (1997).

FIG. 1. Temperature dependence of (a) specific heat (dark cyan curve) and c-axis capacitance (purple curve) of DyMnO$_3$. $T_N$ and $T_S$ denote the antiferromagnetic and spiral phase transitions of Mn$^{3+}$ spins, respectively. (b) Ferroelectric polarization $P_c$ along c-axis when poled with ± 185 V (~ ± 487 kV/m) from $T$ = 40 K to 2 K. (c) Pyroelectric current and (d) the integrated $P_c$ as a function of temperature after being poled with ± 185 V from $T$ = 100 K to 2 K.

FIG. 2. (a) Temperature profiles of ferroelectric polarization $P_c$ with different poling voltages applied from $T$ = 100 K to 2 K. (b) $P_c$ as a function of temperature with 185 V voltage applied from various temperatures $T_{pole}$ to 2 K. Inset shows $P_c$ at 30 K extracted from Fig. 2(b). (c) Pyroelectric current along the c-axis measured after poling the sample from $T$ = 25 K to 2 K with ± 185 V voltages; top and bottom insets show a full view and an expanded view of the corresponding pyroelectric current. (d) Thermally stimulated depolarization current (TSDC) measured under different warming rates (b = dT/dt) after poling the sample from $T$ = 100 K to 50 K with 185 V applied voltage. Inset shows the fitting using an equation $\ln \frac{T_m^2}{b} = \frac{E}{k_B T_m} + \ln \frac{\tau_0 E}{k_B}$, which describes the relation between $T_m$ and b, where $T_m$ represents the temperature at which the TSDC shows a maximum. $E$ is activation energy, $k_B$ is Boltzmann constant, and $\tau_0$ is the relaxation at high temperature limit.[28] The extracted activation energy is $E$ = 0.158 eV.

FIG. 3. (a) Polarization $P_c$ measured after poling the sample from $T$ = 100 K to 25 K with V$_p$s, then zero-field-cooling (ZEC) it to 2 K. (b) Polarization difference between 7 K and 20 K ($P_c^{6.5K}$ - $P_c^{20K}$), i.e., the net polarization induced in the FE state as a function of HTP polarization at 20 K extracted from (a). (c) $P_c$ measured after poling the sample from $T$ = 100 K to 25 K with 185 V, then from 25 K to 2 K with various V$_p$s applied. (d) Polarization difference ($P_c^{6.5K}$ - $P_c^{20K}$) extracted from (c) as a function of voltage applied from $T$ = 25 K to 2 K.

FIG. 4. Schematics showing various configurations of spin chirality and low temperature FE states under different cooling procedures: (a) Poling the sample from $T$ = 25 K to 2 K with V$_p$ = 185 V; (b) Poling the sample from $T$ = 100 K to 25 K with V$_p$ = 185 V, then zero-field-cooled (ZEC) down to 2 K without electric field; (c) Poling the sample from $T = 100$ K to 25 K with $V_p = 185$ V, then cooled down to 2 K with $V_p = 30$ V; (d) Poling the sample from $T = 100$ K to 2 K with $V_p = 185$V.

FIG. 1.

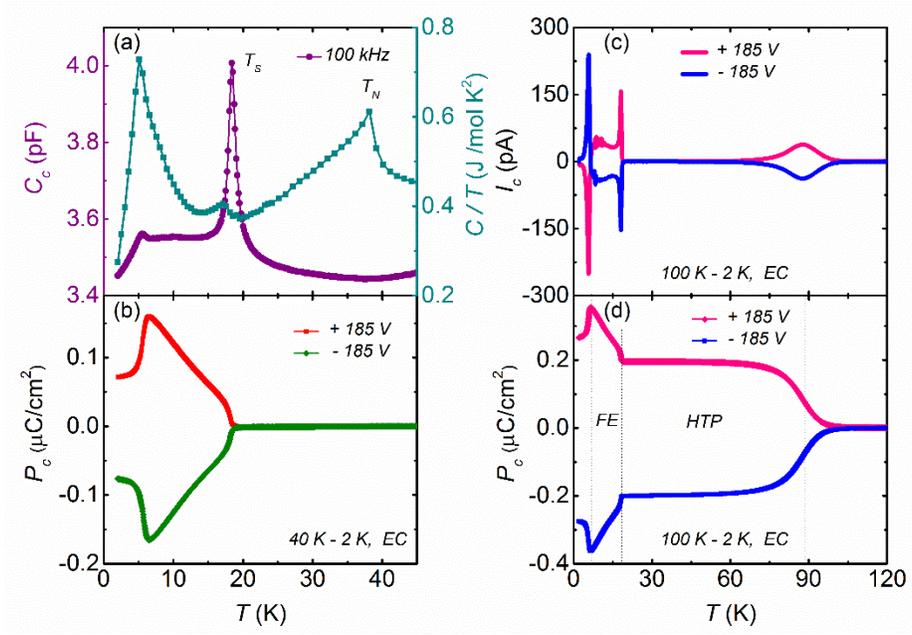

FIG. 2.

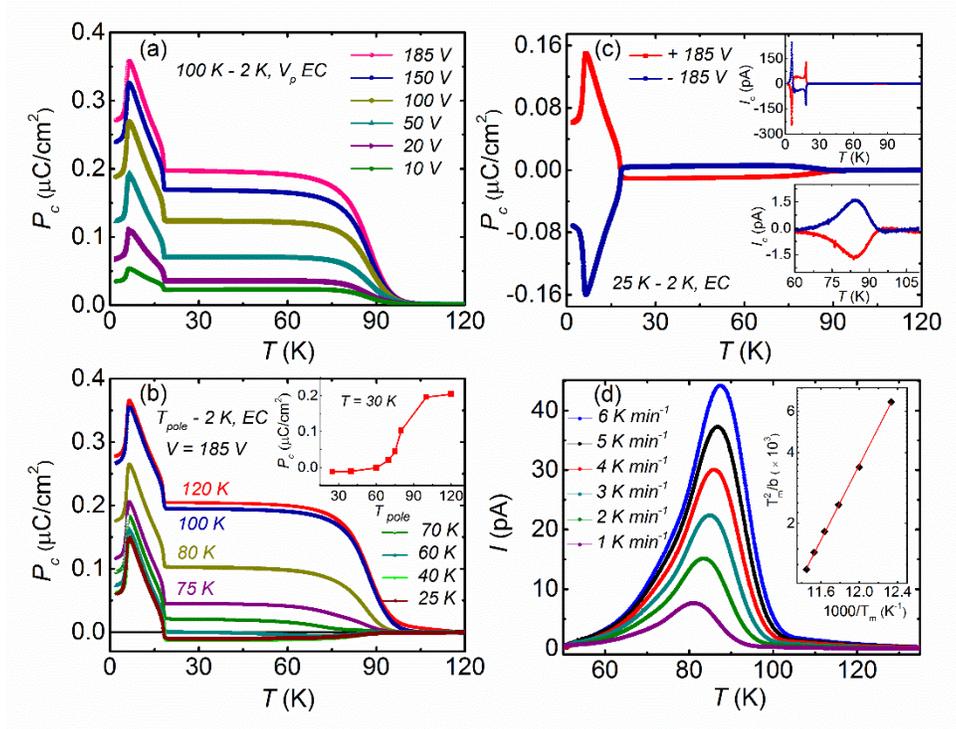

FIG. 3.

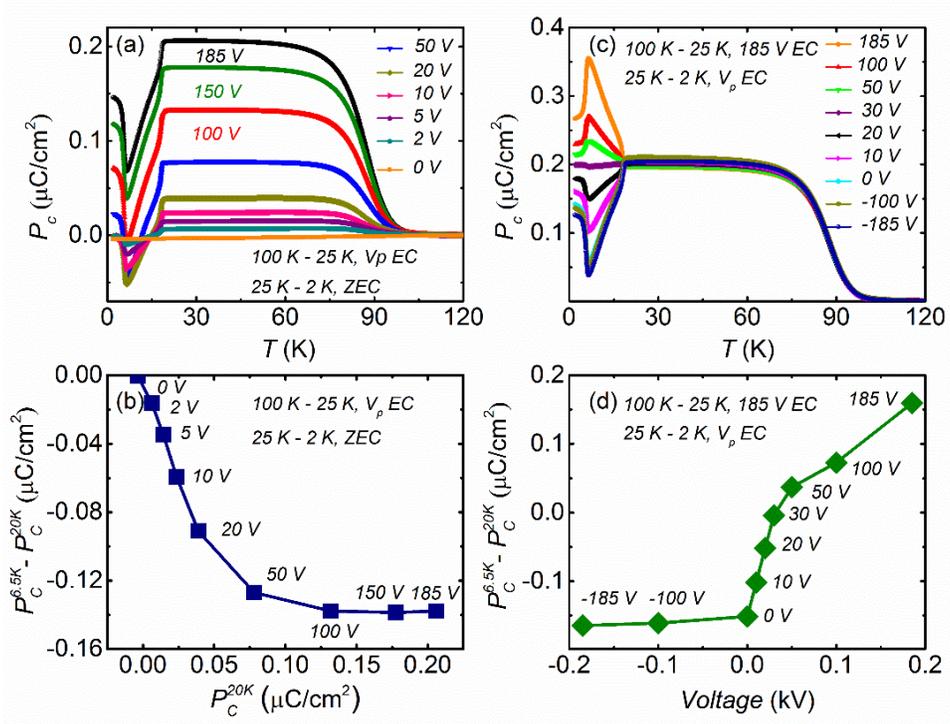

FIG. 4.

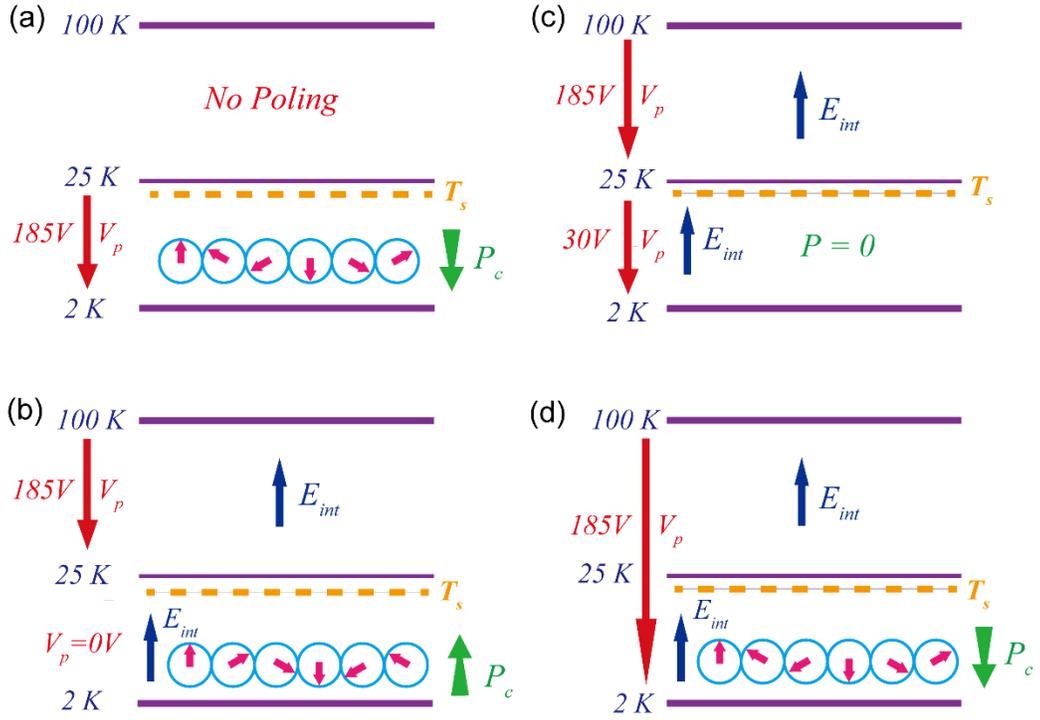